\RequirePackage{fix-cm}
\documentclass[twocolumn,epjc3]{svjour3}  
\smartqed  % flush right qed marks, e.g. at end of proof
\RequirePackage{graphicx}
\renewcommand\makeheadbox{{%
\hbox to0pt{\vbox{\baselineskip=10dd\hrule\hbox
to\hsize{\vrule\kern3pt\vbox{\kern3pt
\hbox{Fermilab-PUB-15-043-T}
\kern3pt}\hfil\kern3pt\vrule}\hrule}%
\hss}}}
\journalname{Eur. Phys. J. C}
\begin{document}

\title{A Multi-Threaded Version of MCFM}


\author{John~M.~ Campbell\thanksref{e1,addr1}
        \and
        R. ~Keith~Ellis\thanksref{e2,addr1} %etc.
        \and
         Walter~T.~Giele\thanksref{e3,addr1} %etc.}
}

%Grants or other notes
\thankstext{e1}{e-mail: johnmc@fnal.gov}
\thankstext{e2}{e-mail: ellis@fnal.gov}
\thankstext{e3}{e-mail: giele@fnal.gov}


\institute{Fermilab, PO Box 500, Batavia, IL 60510, USA \label{addr1}
}

\date{\today}

\maketitle

\begin{abstract}

We report on our findings modifying MCFM using OpenMP to implement
multi-threading. By using OpenMP, the modified MCFM will execute on any 
processor, automatically adjusting to the number of available threads. 

We modified the integration routine VEGAS to distribute the event evaluation over
the threads, while combining all events at the end of every iteration
to optimize the numerical integration.

Special care has been taken that the results of the Monte Carlo integration 
are independent of the number of threads used, to facilitate the validation
of the OpenMP version of MCFM.

%\keywords{Event generation \and parallel programming}
\end{abstract}

\section{Overview}

An important aspect of Monte Carlo programs is evaluation speed
and ease of use. A faster overall evaluation speed not only means
that more complicated processes can be evaluated, but it also allows for
more experimentation as results are returned in a shorter time.

Computer processors are increasing their computational power by
including more and more computing cores. 
It is therefore essential for Monte Carlo event generators
to explore the possibility of a parallel implementation of the code
by taking advantage of the multiple threads to reduce the evaluation time for a given number of events.
By properly implementing the use of multi-threading, the Monte Carlo
evaluation speed will
scale with the number of cores; this process will continue 
as more and more cores become available in the future. 
Monte Carlo event generators are well suited to take advantage
of multi-core processors. 
Parallelization is straightforward as each generated event 
is evaluated independently, while the
results of these evaluations are all combined to optimize the
numerical integration.

The reason processors increase the number of cores instead of the
processor frequency is the limitation
deriving from the growth of the power consumption of the chip.
The power consumption in a chip is given by the equation
\begin{equation}
    P = C V^2 f
\end{equation}
where $P$ is power, $C$ is the capacitance being switched per clock cycle, 
$V$ is voltage, and $f$ is the processor frequency 
(cycles per second). As the clock speed increases the power (and hence heat) 
grows linearly. By having two circuits in 
parallel, we can double the capacitance and halve the clock speed. 
The voltage determines the rate at which the capacitance
charges and discharges, so that a slower clock speed can run with lower voltages.
At half the clock speed, we can approximately halve the voltage, 
leading to a saving in power
without a compromise in performance.
The use of many cores in this fashion may allow the growth of
computing power to continue following Moore's law in the future.
It is therefore imperative that software evolve to take advantage
of these developments.

Currently the Intel Xeon-Phi coprocessor with 240 processor threads 
and General Purpose Graphics Processing Units (GPGPU's) with 
up to 2,880 gpu cores are the most extreme implementation 
of this approach to increasing the computational power. 
The Xeon Phi is the first generation of the Intel MIC (Many
Integrated Cores) hardware. With an improved version
of this coprocessor planned for release in the summer of 2015, 
further speed-ups can be expected.

We will explore using this co-processor and more conventional processors using OpenMP.
Specifically, we will test our OpenMP version of MCFM\footnote{MCFM-7.0 which runs under the OpenMP protocol as described in this paper can be downloaded from the mcfm.fnal.gov website.} on an Intel
Core I7-4770 (4 hardware threads),
a dual Intel Xeon X5650 (2x6 hardware threads), 
a quadruple AMD 6128 HE Opteron (4x8 hardware threads) and the
Intel Xeon-Phi 5110P (240 hardware threads). 
Note that the Intel Core i7 comes with 8 hyperthreads, 
2 software threads per core. The core can execute only one of the threads and quickly switch 
to the other thread if the current thread is waiting.
As we will see this is of limited benefit for our application.

The OpenMP standard\footnote{
`OpenMP (Open Multi-Processing) is an API that supports
multi-platform shared memory multiprocessing programming in C, C++,
and Fortran, on most processor architectures and operating systems,
including Solaris, AIX, HP-UX, Linux, Mac OS X, and Windows
platforms. It consists of a set of compiler directives, library
routines, and environment variables that influence run-time behavior.
OpenMP is managed by the nonprofit technology consortium
OpenMP Architecture Review Board (or OpenMP ARB), jointly defined by a
group of major computer hardware and software vendors, including AMD,
IBM, Intel, Cray, HP, Fujitsu, Nvidia, NEC, Red Hat, Texas
Instruments, Oracle Corporation, and more.', from Wikipedia.}
\cite{OpenMP} is a good choice for implementing
parallel programming. It is native to both the Intel and GNU compilers
and can be invoked by including the `openmp'-flag during compilation.
No special libraries or other software need to be installed.
The OpenMP compiler directives are simply implemented 
as comment statements in either FORTRAN
or C/C++ code. This has the advantage that the code can be compiled 
without the `openmp'-flag. In this case the OpenMP directives are interpreted as
comments by the compiler. Furthermore, we can implement the parallelism 
with only minor alterations to the original code by just adding these compiler directives.

The further layout of our paper is as follows. 
In section 2 we discuss some details and considerations for implementing OpenMP 
into the FORTRAN code of MCFM~\cite{Campbell:2010ff,Campbell:2011bn} (similar 
considerations will hold for C/C++ code).
The numerical performance of the parallel code is explored in section 3
using several different processors.
Finally, in section 4 we sum up our conclusions and review further possible developments
for the OpenMP MCFM program.

\section{Implementing OpenMP in MCFM}

\subsection{MCFM}
%Easiest way to parallelize: Vegas (common grid to generate random numbers, threaded event 
%generation and matrix element evaluation) 

MCFM is a parton level integrator, developed over many years at
Fermilab, that calculates cross sections and distributions of
kinematic variables for hard scattering processes in hadron-hadron
collisions~\cite{Campbell:2010ff}. More than $300$ processes are included, the majority of them
calculated at next-to-leading order in the strong coupling. The event
generator consists of an adaptive integration routine which generates
the events.  The returned event probabilities are used to further
optimize the integration using importance sampling. 
The program spends the bulk of its time in
the event evaluation routines.

For MCFM the multi-dimensional integration is implemented using
VEGAS~\cite{Lepage:1977sw,Press:1992zz}. 
It produces several iterations of sets of events. After each
iteration the grid is optimized to reduce the weight fluctuations in
the integration so that faster convergence is obtained.
This offers an obvious and straightforward way to parallelize the
program.
While the grid optimization is not parallelized, so that all
the results can be combined, the individual generation of phase space points and
subsequent matrix element evaluation can be done in parallel as no
data sharing is required between different events. This allows
the parallel program to access all evaluated events to obtain maximum
convergence, while the event evaluation is sped up considerably by using each thread for a
different event generation and evaluation.

This should be contrasted with simultaneous running of an
individual program on each thread. 
In this case the grids in each program are only updated with
the events from that particular thread, leading to a worse convergence. 
The parallel version offers the advantage of combining the events from
all threads for the grid optimization.

\subsection{OpenMP-MCFM}
Here we detail the work needed to produce an OpenMP implementation of MCFM.
The MCFM code is large and complicated. To convert MCFM to 
an OpenMP supported MCFM requires some thought and work. We used as
goals (a) to minimize the changes in the original code and (b) to implement
the parallelism through comment compiler directives as much as
possible. This makes the code compilable with or without the OpenMP flag.
Another goal (c) was to make sure the program generates the same events
independent of the number of threads used. We verified that the 
results obtained are independent of the number of threads used to evaluate
the cross sections. This greatly helps to validate that
the implementation of the parallel code is correct. 

Almost all the work to be done is to make sure variables are correctly
assigned. In a parallel program we have to decide whether a variable
is  global (i.e. potentially shared by threads) or local to the
thread (i.e. each thread has its own version of the variable).

%How to convert a fortran MC to an OpenMP fortran MC:
%Data structures In parallel region:
%1. Data statements variables must be in save statement
%2. Variables in save statements must be threadprivate. No automatic save flag 
%3.  Common blocks whose variables are changed in the parallel region 
%must be declared threadprivate each time the common block is declared 
%4. Common blocks whose variables are defined/changed outside the 
%parallel region in addition to being changed inside the parallel 
%region neet to be threadprivate and need to be included in the copyin 
%statement at the start of the parallel region 

The most labor-intensive part is the treatment of data structures. The
following rules will lead to a successful parallelization. For all the
code running in parallel one has to implement the following steps:
\begin{itemize}
\item All variables in DATA statements in the parallel region
have to be included in SAVE
statements ensuring they are declared for each thread. If not done, the
variables are not necessarily initialized.
\item All variables in SAVE statements in the parallel region
  must be made `thread private' in the respective functions and
  subroutines.
\item All common blocks whose variables are defined or changed in the
  parallel region have to
  be declared `thread private' each time the common block is declared.
\item All common blocks whose variables are defined or changed outside
  the parallel region in addition to being changed in the parallel
  region need to be declared `thread private'. 
  To ensure the values are copied to each thread at the
  start of the parallel region a COPYIN directive including the
  common block has to be issued.
\end{itemize}
Note that, where necessary, variables and common blocks are made `thread private'
by adding the THREADPRIVATE directive to the function or subroutine~\cite{OpenMP}.

The MCFM code was originally written in FORTRAN 77, but parts of the code 
now require a FORTRAN 90 compiler. In view of the special treatment required for data statements,
indicated above, it is beneficial to eliminate data statements wherever
they are not needed. FORTRAN 90 allows parameter arrays, so it is
useful to replace the FORTRAN 77 legacy data arrays by parameter arrays wherever possible. 
%Treatment of random numbers and getting identical events independent  
%on number of threads used  
 
To ensure that the same events are generated,
independent of the number of threads used, we
have to ensure VEGAS generates the same sequence of groups of pseudo-random
numbers used to generate the momenta in an event. To do this we use the CRITICAL
directive forcing the pseudo-random number generator to run
serially, when assigning the groups of pseudo-random numbers to a thread. When
looking at all threads combined, the same groups of random
numbers will be generated, and consequently the same set of
events. The order in which the groups of random numbers 
are accessed by the threads is not identical and varies from
run to run (i.e.\ which thread reaches the critical region first) but in the
end the same events are always generated. A named CRITICAL directive provides 
a way of distinguishing CRITICAL regions in different parts of the program.
When a thread arrives at a CRITICAL directive, it waits until no other thread
is executing a critical region with the same name.
 
The ATOMIC construct, which applies only to the specific assignment statement that follows
it, can be an efficient alternative to a CRITICAL region. The statement following an ATOMIC 
directive is executed by all threads, but only one thread at a time can execute the statement.
%Kahan summation needed to get identical cross sections

This is still not sufficient to reach identical results for the cross
section. The reason for this is numerical rounding differences due to
the fact that the resulting weights are added in different orders. 
Using Kahan summation~\cite{Kahan} will ameliorate rounding error, leading
not only to identical cross section results but also to more accurate results.

We checked that all processes in MCFM produce identical results independent 
of the number of threads and in agreement with the non-parallel version of
MCFM (version 6.8).
%At this time we did not implement the APPLgrid option in the OpenMP version.

\section{Performance of OpenMP-MCFM}

\subsection{Runtime considerations}

We used version 3.0 of OpenMP to prepare our code, which includes
all of the compiler directives discussed above.
To compile the program the `openmp'-flag has to be included. The
resulting executable will use by default all available threads during
execution. Note that if the program is compiled without the OpenMP
flag it will not use multi-threading.  To lower the number of
threads used, two options are available. The first option uses the
environmental variable OMP\_NUM\_THREADS. This variable
can be set to the number of
threads the OpenMP executable will use. Another possibility is to
include the omp\_lib.h library in the program which gives access
to in-program OpenMP commands. The function call 
omp\_set\_num\_threads(int) sets the number of threads used
to the value of the integer `int'.
This allows
for a dynamical change of the number of threads during execution. The
library also gives access to many more OpenMP function calls, that are currently 
of no importance in running MCFM.

Another consideration is the memory stack size to be used by each thread. The
default size of the stack is not specified by the OpenMP standard. If the stacksize is too
small the program will crash with a segmentation fault or other unexpected behaviour. 
To be able to execute
all processes in MCFM the stack size should be set to 16,000 or higher using 
the  environmental variable OMP\_STACKSIZE (though for most processes in MCFM a much 
smaller stacksize suffices).

\subsection{Results}

\begin{table}\begin{center}
\begin{tabular}{|c||rrr||r||c|}
\hline 
\multicolumn{6}{|c|}{Intel Core I7-4770}\\
\hline 
\multicolumn{1}{|c||}{Thr.} &\multicolumn{3}{c||}{Time (sec)}
&\multicolumn{1}{c||}{Acc.} & \multicolumn{1}{c|}{Eff.} \\ 
 & min & avg & max & avg &\multicolumn{1}{c|}{(\%)} \\
\hline
1 & 1.67 & 1.69 & 1.70 & 1.00 & 100.00\\
2 & 0.83 & 0.83 & 0.83 & 2.02 & 101.21\\
3 & 0.57 & 0.57 & 0.58 & 2.94 &   97.88\\
4 & 0.44 & 0.44 & 0.44 & 3.80 &   95.12\\
5 & 0.40 & 0.40 & 0.40 & 4.18 &   83.50\\
6 & 0.37 & 0.37 & 0.37 & 4.55 &   75.78\\
7 & 0.34 & 0.34 & 0.34 & 4.92 &   70.26\\
8 & 0.32 & 0.32 & 0.32 & 5.25 &   65.65\\
\hline 
\end{tabular}
\caption[]{}
%\end{center}\end{table}

%\begin{table}\begin{center}
\begin{tabular}{|c||rrr||r||c|}
\hline 
\multicolumn{6}{|c|}{Dual Intel Xeon X5650}\\
\hline 
\multicolumn{1}{|c||}{Thr.} &\multicolumn{3}{c||}{Time (sec)}
&\multicolumn{1}{c||}{Acc.} & \multicolumn{1}{c|}{Eff.} \\ 
 & min & avg & max & avg &\multicolumn{1}{c|}{(\%)} \\
\hline
  1 & 2.88 & 2.89 & 2.89 &   1.00 & 100.00\\
  2 & 1.49 & 1.49 & 1.50 &   1.94 &   96.76\\
  3 & 0.99 & 1.00 & 1.00 &   2.90 &   96.60\\
  4 & 0.75 & 0.75 & 0.75 &   3.85 &   96.13\\
  6 & 0.50 & 0.50 & 0.51 &   5.72 &   95.30\\
  8 & 0.38 & 0.38 & 0.38 &   7.57 &   94.59\\
10 & 0.31 & 0.31 & 0.31 &   9.37 &   93.66\\
12 & 0.26 & 0.26 & 0.26 & 11.16 &   92.96\\
\hline 
\end{tabular}
\caption[]{}
%\end{center}\end{table}

%\begin{table}\begin{center}
\begin{tabular}{|c||rrr||r||c|}
\hline 
\multicolumn{6}{|c|}{Quadruple AMD 6128 HE Opteron}\\
\hline 
\multicolumn{1}{|c||}{Thr.}  &\multicolumn{3}{c||}{Time (sec)}
&\multicolumn{1}{c||}{Acc.} & \multicolumn{1}{c|}{Eff.} \\ 
& min & avg & max & avg  &\multicolumn{1}{c|}{(\%)} \\
\hline
  1 & 3.79 & 3.80 & 3.80 &   1.00 & 100.00\\
  2 & 2.00 & 2.02 & 2.05 &   1.88 &   94.06\\
  3 & 1.36 & 1.37 & 1.38 &   2.77 &   92.42\\
  4 & 1.03 & 1.04 & 1.05 &   3.66 &   91.52\\
  8 & 0.54 & 0.54 & 0.54 &   7.00 &   87.44\\
12 & 0.38 & 0.38 & 0.38 &   9.98 &   83.13\\
16 & 0.33 & 0.33 & 0.33 & 11.44 &   71.52\\
32 & 0.83 & 0.84 & 0.86 &   4.50 &   14.06\\
\hline 
\end{tabular}
\caption[]{}
%\end{center}\end{table}

%\begin{table}\begin{center}
\begin{tabular}{|c||rrr||r||c|}
\hline 
\multicolumn{6}{|c|}{Intel Xeon Phi 5110P}\\
\hline 
\multicolumn{1}{|c||}{Thr.} &\multicolumn{3}{c||}{Time (sec)}
&\multicolumn{1}{c||}{Acc.} & \multicolumn{1}{c|}{Eff.} \\ 
 & min & avg & max & avg &\multicolumn{1}{c|}{(\%)} \\
\hline
    1 & 23.09 & 23.12 & 23.15 &   1.00 & 100.00\\
    2 & 12.10 & 12.12 & 12.14 &   1.91 &   95.39\\
    3 &   8.14 &   8.22 &   8.53 &   2.81 &   93.78\\
    4 &   6.16 &   6.21 &   6.38 &   3.72 &   93.11\\
  16 &   1.66 &   1.67 &   1.68 & 13.86 &   86.61\\
  32 &   1.39 &   1.39 &   1.40 & 16.61 &   51.89\\
  64 &   1.41 &   1.41 &   1.41 & 16.39 &   25.61\\
128 &   1.44 &   1.44 &   1.45 & 16.02 &   12.52\\
240 &   1.52 &   1.52 &   1.53 & 15.19 &     6.33\\
\hline 
\end{tabular}
\end{center}
\caption{}{The LO evaluation of $PP\rightarrow H(\rightarrow b\bar b) + 2$
  jets using 4x1,000+10x10,000 Vegas events for the 4 different hardware configurations.}
\end{table}

\begin{table}\begin{center}
\begin{tabular}{|c||rrr||r||c|}
\hline 
\multicolumn{6}{|c|}{Intel Core I7-4770}\\
\hline 
\multicolumn{1}{|c||}{Thr.} &\multicolumn{3}{c||}{Time (sec)}
&\multicolumn{1}{c||}{Acc.} & \multicolumn{1}{c|}{Eff.} \\ 
 & min & avg & max & avg &\multicolumn{1}{c|}{(\%)} \\
\hline
1 & 238.83 & 238.95 & 239.07 & 1.00 & 100.00\\
2 & 120.16 & 120.45 & 120.73 & 1.98 &   99.19\\
3 &   81.99 &   82.03 &   82.07 & 2.91 &   97.10\\
4 &   63.01 &   63.02 &   63.02 & 3.79 &   94.80\\
5 &   58.67 &   58.69 &   58.71 & 4.07 &   81.43\\
6 &   54.84 &   54.85 &   54.86 & 4.36 &   72.61\\
7 &   51.52 &   51.53 &   51.54 & 4.64 &   66.24\\
8 &   48.62 &   48.63 &   48.64 & 4.91 &   61.42\\
\hline 
\end{tabular}
\caption[]{}
%\end{center}\end{table}

%\begin{table}\begin{center}
\begin{tabular}{|c||rrr||r||c|}
\hline 
\multicolumn{6}{|c|}{Dual Intel Xeon X5650}\\
\hline 
\multicolumn{1}{|c||}{Thr.} &\multicolumn{3}{c||}{Time (sec)}
&\multicolumn{1}{c||}{Acc.} & \multicolumn{1}{c|}{Eff.} \\ 
 & min & avg & max & avg &\multicolumn{1}{c|}{(\%)} \\
\hline
  1 & 496.43 & 496.43 & 496.44 &   1.00 & 100.00\\
  2 & 249.73 & 249.83 & 249.94 &   1.99 &   99.35\\
  3 & 166.20 & 166.41 & 166.62 &   2.98 &   99.44\\
  4 & 124.58 & 124.58 & 124.59 &   3.98 &   99.62\\
  6 &   83.01 &   83.04 &   83.06 &   5.98 &   99.64\\
  8 &   62.24 &   62.26 &   62.29 &   7.97 &   99.66\\
10 &   49.79 &   49.80 &   49.80 &   9.97 &   99.69\\
12 &   41.46 &   41.46 &   41.46 & 11.97 &   99.78\\
\hline 
\end{tabular}
\caption[]{}
%\end{center}\end{table}

%\begin{table}\begin{center}
\begin{tabular}{|c||rrr||r||c|}
\hline 
\multicolumn{6}{|c|}{Quadruple AMD 6128 HE Opteron}\\
\hline 
\multicolumn{1}{|c||}{Thr.} &\multicolumn{3}{c||}{Time (sec)}
&\multicolumn{1}{c||}{Acc.} & \multicolumn{1}{c|}{Eff.} \\ 
 & min & avg & max & avg &\multicolumn{1}{c|}{(\%)} \\
\hline
  1 & 806.86 & 806.98 & 807.10 &   1.00 & 100.00\\
  2 & 404.00 & 404.08 & 404.17 &   2.00 &   99.85\\
  3 & 269.26 & 269.37 & 269.48 &   3.00 &   99.86\\
  4 & 201.96 & 201.99 & 202.02 &   4.00 &   99.88\\
  8 & 101.03 & 101.05 & 101.07 &   7.99 &   99.83\\
12 &   67.41 &   67.41 &   67.41 & 11.97 &   99.76\\
16 &   50.56 &   50.56 &   50.56 & 15.96 &   99.75\\
32 &   25.34 &   25.36 &   25.37 & 31.82 &   99.45\\
\hline 
\end{tabular}
\caption[]{}
%\end{center}\end{table}

%\begin{table}\begin{center}
\begin{tabular}{|c||rrr||r||c|}
\hline 
\multicolumn{6}{|c|}{Intel Xeon Phi 5110P}\\
\hline 
\multicolumn{1}{|c||}{Thr.} &\multicolumn{3}{c||}{Time (sec)}
&\multicolumn{1}{c||}{Acc.} & \multicolumn{1}{c|}{Eff.} \\ 
 & min & avg & max & avg &\multicolumn{1}{c|}{(\%)} \\
\hline
    1 & 3784.45 & 3784.45 & 3784.45 &     1.00 & 100.00\\
    2 & 1906.73 & 1906.73 & 1906.73 &     1.98 &   99.24\\
    3 & 1282.26 & 1282.26 & 1282.26 &     2.95 &   98.38\\
    4 &   958.59 &   958.59 &   958.59 &     3.95 &   98.70\\
  16 &   242.66 &   242.66 &   242.66 &   15.60 &   97.47\\
  32 &   121.25 &   121.25 &   121.25 &   31.21 &   97.54\\
  64 &     62.29 &     62.29 &     62.29 &   60.76 &   94.93\\
128 &     41.22 &     41.22 &     41.22 &   91.81 &   71.73\\
240 &     31.82 &     31.82 &     31.82 & 118.94 &   49.56\\
\hline
\end{tabular}
\end{center}
\caption{}{The NLO evaluation of $PP\rightarrow H(\rightarrow b\bar b) + 2$
  jets using 4x1,000+10x10,000 Vegas events for the 4 different hardware configurations.}
\end{table}

\begin{figure}[h]
    \centering
    \includegraphics[width=9cm,height=7cm]{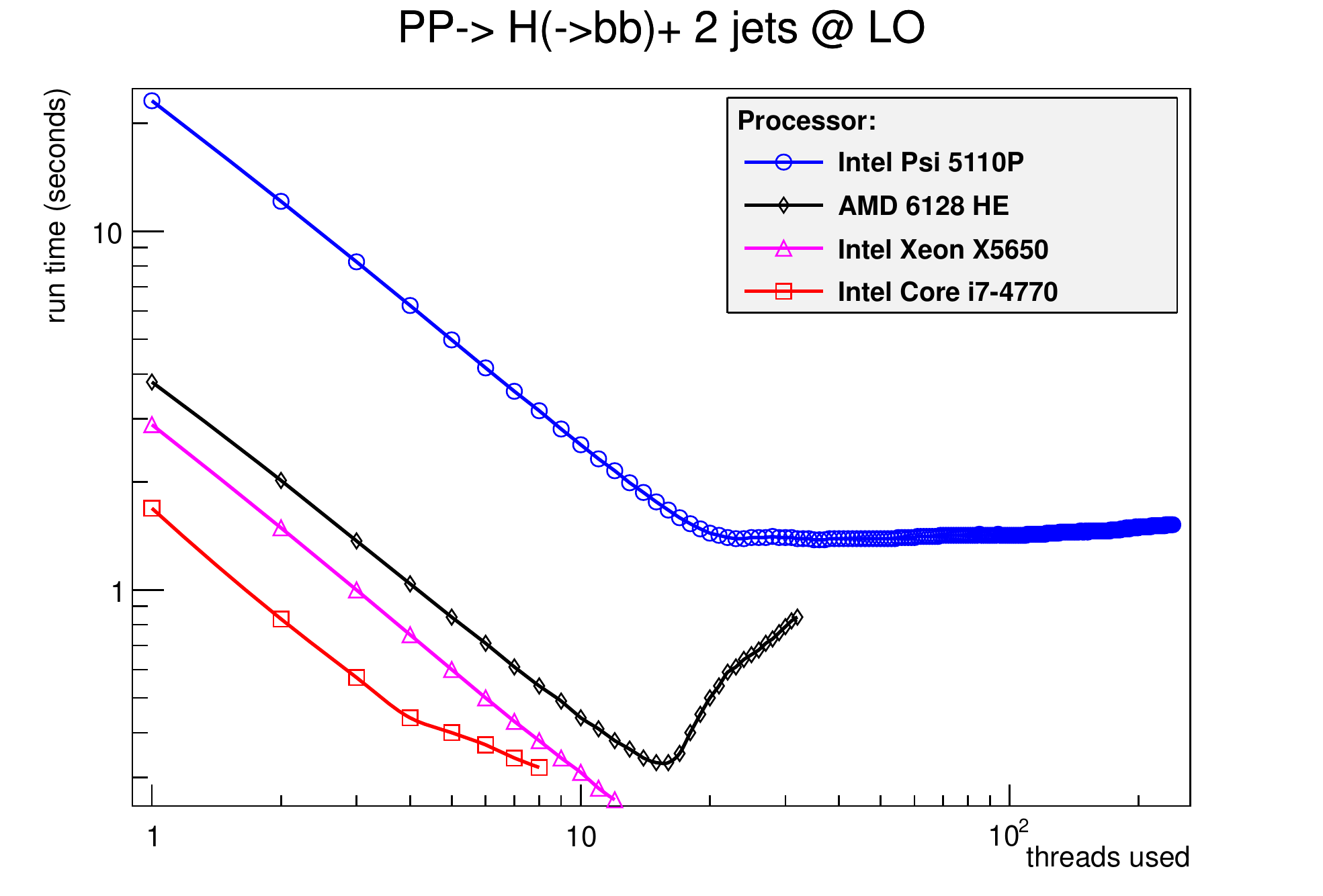}
    \includegraphics[width=9cm,height=7cm]{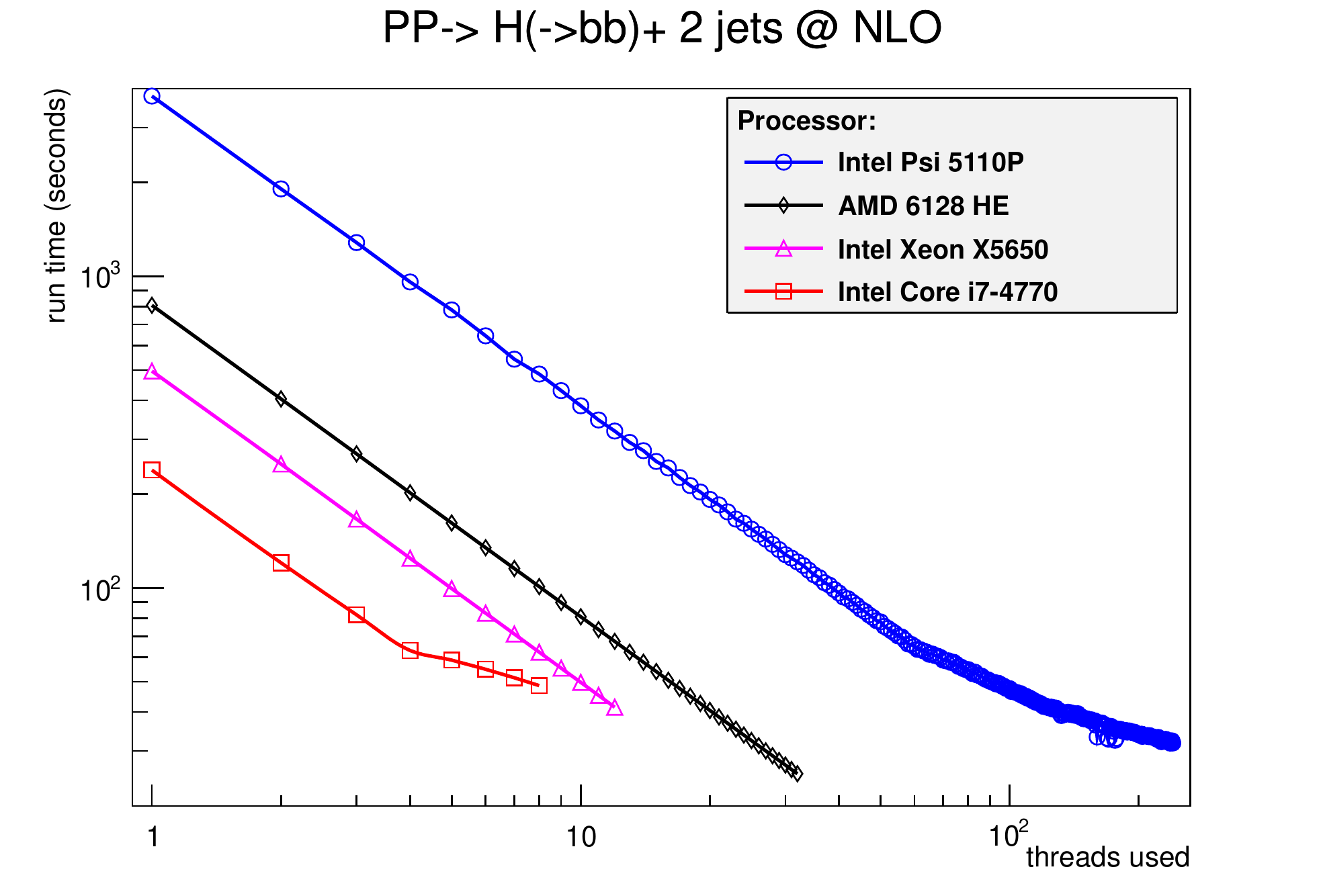}
\caption[]{The evaluation time of $PP\rightarrow H(\rightarrow b\bar
  b) + 2$~jets  using 4x1,000+10x10,000 Vegas events (in seconds) versus the number of
  threads. The top graphs is at LO and the bottom graph at NLO.}
\end{figure}

To benchmark the performance of the parallel version of MCFM we use four
different types of computer hardware. This will test the code on a variety of
hardware configurations with differing clock frequency, number of threads, cache size etc.

The first configuration is a standard desktop with an Intel Core
i7-4770. This processor has 4 cores, each with 2 hyperthreads.  The
second configuration contains two Intel Xeon X5650 processors, each
with 6 cores for a total of 12 cores.  The third configuration
contains four AMD 6128 HE Opteron processors, each with 8 cores for a
total of 32 cores.  The final configuration is an Intel Phi 5110P
coprocessor card connected to a PCI slot.  This coprocessor has 60
cores, each with 4 hardware threads for a total of 240 threads.

While we have validated all processes in this version of MCFM, we pick
one process in particular to study the speedups gained by using
multiple threads. The process we choose is $PP\rightarrow
H(\rightarrow b\bar b) + 2$ jets which describes the production of a
Higgs boson in association with two jets through an effective
gluon-gluon-Higgs vertex. The Higgs boson subsequently undergoes a
two-body decay to two $b$-quarks. Thus the process can have as many as
4 (5) jets in LO (NLO), two of which can come from the Higgs decay.
In lowest order a process with $n$ particles in the final state
requires $3n-4$ phase space integrations and two integrals over parton
density longitudinal fractions.  Thus for this leading order (LO)
process, a 10-dimensional integration is required.  The
next-to-leading (NLO) process requires a $13$-dimensional integration.
The results are contained in Tables 1 through 4 for the LO runs and in
Table 5 through 8 for the NLO runs. The tables contain, for each
configuration and as a function of the number of threads used, the
minimum, average and maximum runtime (in seconds), averaged over 10
runs for the first 3 configurations and 2 runs for the
coprocessor. The acceleration compares the runtime to the single
thread run time by taking the ratio of the two.  Finally we give the
efficiency in percentages. The efficiency is simply the acceleration
divided by the number of threads. For a perfect parallelization,
doubling the number of threads should double the acceleration, leaving
the efficiency at 100\%.  All the average runtime results of the
tables are represented graphically in figure 1 where we plot on a
log-log scale the runtime as a function of the number of threads
used. Note that we do not generate histograms during these
benchmarking runs.

We will first look at the results for the Intel Core i7 in Tables 1
and 5 (and Figure 1).  As we can see the speed-up as far as 4 threads
is good, with an acceleration for LO up to 3.80 and for NLO up to
3.79. As the processor has 4 cores, each thread runs on a different
core. If we use more than 4 threads some or all of the threads will
share a single core with another thread. If one of the threads has to
wait for a memory fetch, the core will switch to the other thread and
start executing. As can be seen, this results in a much slower
speed-up though some speedup is still achieved (from 3.80 for 4
threads to 5.25 for 8 threads at LO and from 3.79 to 4.91 at
NLO). Yet, by using multi-threading on this basic configuration one
can generate around 6.3 million Vegas events at NLO order per hour.
Note that this depends on the cuts applied, as this will affect the
number of rejected events. However, the comparison to other
configurations is illuminating.

The next configuration to consider is the dual socketed X5650
processors giving a total of 12 cores.  The results of Table 2 and 6
show good scaling for LO with a maximum acceleration of 11.16.  At NLO
the acceleration is nearly perfect with a maximum acceleration of
11.97 using 12 threads. The difference in speed-up between LO and NLO
can be understood by the fact that the NLO process is computationally
bound (i.e. the runtime is predominantly determined by floating point
operations), while at LO the computational component is much smaller
and the memory fetch time will become more dominant, i.e. LO is more
bandwidth bound. In other words at LO we do not give the cores enough
floating point operations to keep them fully occupied.  While this
processor runs a factor of 0.56 slower than the Core i7, in the end it
wins out due to the use of 12 cores.  By using multi-threading on this
configuration one can generate 8.7 million Vegas events at NLO order
per hour.

In Tables 3 and 7 we move on to the quad-socketed AMD 6128 HE Opteron
processors giving a total of 32 cores. We very clearly see the effect
of the bandwidth bound LO and the computational bound NLO. The NLO
gives nearly perfect acceleration of 31.82, while LO reaches its
maximum acceleration of 11.44 using 16 of the 32 cores. Using more
than 16 cores actually makes the evaluation time slower as the
bandwidth limitation becomes more important than the computational
one.  Despite being slower by a factor 0.35 compared to the Core I7
processor one can generate 14.2 million Vegas events at NLO per hour.

The final configuration is the Xeon-Phi coprocessor with 240 hardware
cores. To achieve good acceleration it is crucial to have a
computational bound calculation.  This is dramatically demonstrated in
Tables 6 and 8. At LO it achieves its fastest evaluation time using
around 32 threads with an acceleration around 16.61. However at NLO
the coprocessor keeps accelerating up to 240 threads for the
evaluation time of 31.82 seconds, giving a maximum acceleration of
around 119.  One can generate 10.7 million Vegas events at NLO per
hour.  While this co-processor has an impressive acceleration of over
a factor of 100, the processing speed of a single core is slow. (It is
a factor of 0.07 slower than the Core i7). The next iteration of the
co-processor is expected to be significantly faster, making this MIC
architecture very attractive in the near future.  It is worth noting
this co-processor is a PCI-bus card which, given the right
configuration, can be added to a desktop turning it into a very
powerful stand-alone event generator.

\begin{figure}[h]
    \centering
    \includegraphics[width=9cm,height=7cm]{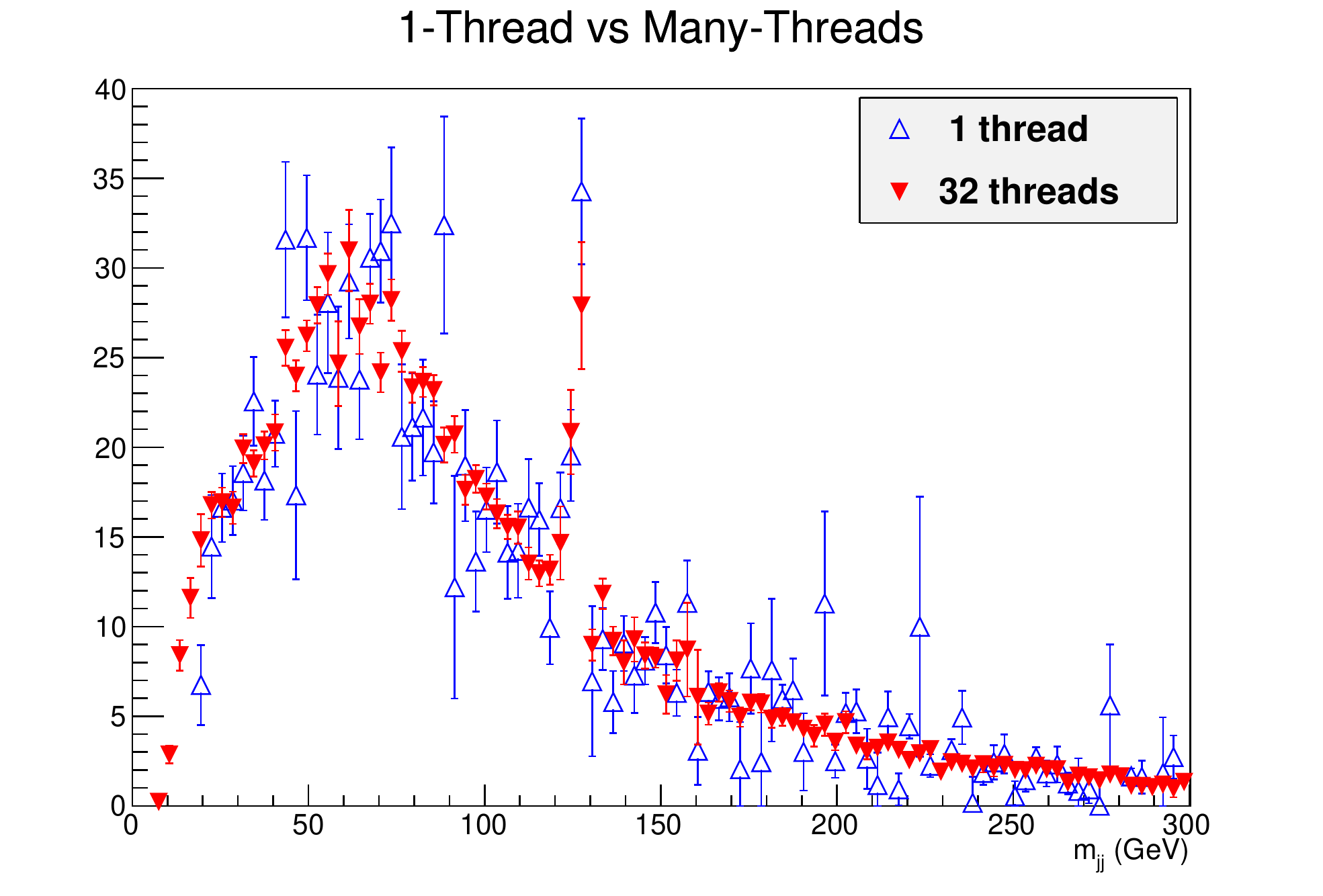}
\caption[]{The di-jet differential cross section for $PP\rightarrow H(\rightarrow b\bar
  b) + 2$~jets  at NLO using 1 hour of running time on the Intel Core
  I7-4770 using a single thread and on the quadruple AMD 6128 HE
  Opteron using all 32 threads. The peak at $m_{jj}=125$~GeV when the
  two jets come from the decay of the Higgs boson is visible.}
\end{figure}
\begin{figure}[h]
    \centering
    \includegraphics[width=9cm,height=7cm]{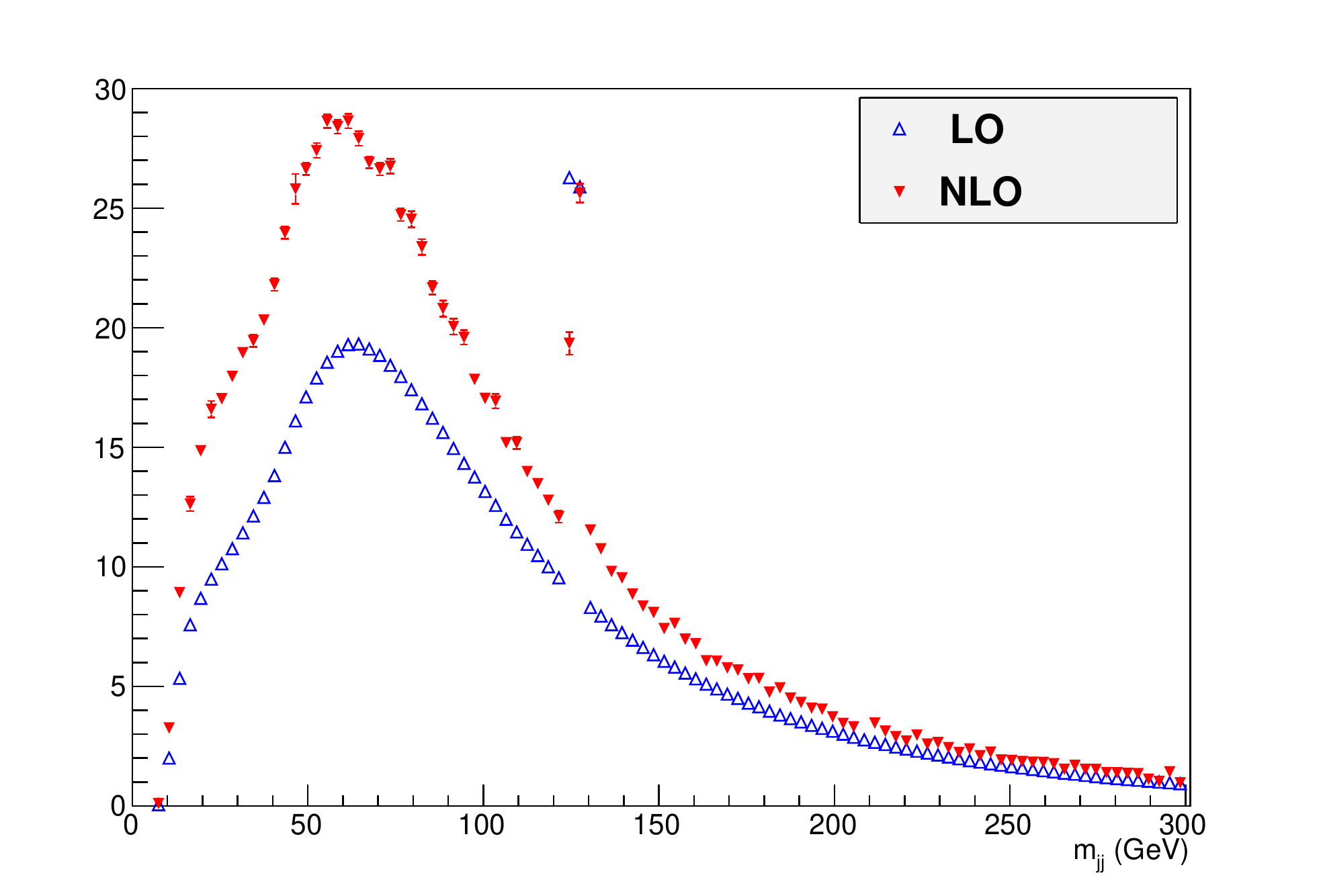}
\caption[]{The di-jet differential cross section for $PP\rightarrow H(\rightarrow b\bar
  b) + 2$~jets using 4x1,500,000+10x15,000,000 events. At LO we use
the Dual Intel Xeon X5650 with 12 threads (about 12 minutes of runtime) and at NLO we use the
quadruple AMD 6128 HE Opteron (about 22 hours of runtime ) with 32 threads. The peak when the
two jets come from the decay of the Higgs boson is clearly visible.}
\end{figure}
 
To see the impact of the faster running we show in Figure 2 the
results for the di-jet mass invariant mass distribution. We compare the fastest single thread
configuration  (the Core-I7) and the fastest multi-thread
configuration (quad AMD)  using approximately 1 hour of runtime for
each. We see that the single thread run is insufficient for any useful
exploratory runs. In contrast one hour of running on the multi-threaded
system gives a good result.
Finally, in Figure 3 we make the di-jet distribution using about 24 hours of
runtime which is more than sufficient to produce a stable final result. 

\section{Conclusions}

To conclude we see that the threaded version of MCFM accelerates well on different architectures.
The computationally bound NLO processes scale well with the number of threads 
and the evaluation speeds are significantly improved. In particular, the performance of the Xeon-Phi
coprocessor is impressive. A new coprocessor is to be released in the summer of 2015, promising
even faster evaluation times. Moreover, this new version will also be available in a socketed
version, removing the PCI-bus and hopefully alleviating the bandwith-bound issues of LO.
This will make the Xeon-Phi coprocessor a very attractive option for
Monte Carlo generators in the near future. 

As we have shown, we have successfully implemented a parallel
version of MCFM.  It instantly reduces the execution time
dependent on the hardware configuration of the system (i.e.
number of cores, cache configuration, memory bandwidth, clock frequency etc)
without any intervention of the user of MCFM.
For the computing intensive next-to-leading
order processes we obtain very good accelerations on all processors. 
In particular, utilizing the Xeon-Phi coprocessor with 240 hardware cores
yields an acceleration of order 100 over running on a single thread.

The new Xeon-Phi processor, to be released in summer 2015, will
overcome most of the bandwith limitation  to which the compute-light
leading order processes are subject. Moreover the new processor will be
substantially more powerful, giving us accelerations well over a
factor of 100.  Now that we have improved the speed of MCFM, we can
implement more complicated processes in the event generator and still
get acceptable evaluation times. Possibilities could
include adding more jets to current processes in MCFM or proceeding to
next-to-next-to leading order processes.

\section*{Acknowledgments}
The numerical work on the Intel Xeon-Phi processor was performed using
the Fermilab MIC development cluster funded by the DOE Office of
Science and operated by the Fermilab scientific computing HPC
department.  
We acknowledge useful discussions with Don Holmgren and
James Simone.  This research is supported by the US DOE under contract
DE-AC02-07CH11359.

%\appendix
%\section{Code examples}

\bibliographystyle{spphys}       % APS-like style for physics
\bibliography{mcfm_omp}   % name your BibTeX data base

\end{document}